# Has the COVID-19 Pandemic Altered the Traditional View about Women's Active Work?

Eiji YAMAMURA *

Department of Economics, Seinan Gakuin University/ 6-2-92 Nishijin Sawaraku Fukuoka, 814-8511.

*Corresponding Author)

Email: yamaei@seinan-gu.ac.jp

Fumio OHTAKE

Center for Infectious Disease Education and Research, Osaka University, Japan,

Email:  ohtake@cider.osaka-u.ac.jp

**Declarations**

**Abstract**

This study investigates how the view about women's active work changed after the outbreak of the novel coronavirus 2019 (COVID-19) disease. We use individual-level panel data from 2016 to 2024 that cover the period before and after the pandemic. The major findings are as follows: (1) men were more likely to have a positive view than women before COVID-19, whereas women became more likely to have a positive view compared to men after COVID-19; (2) both of men and women were more likely to have a positive view after COVID-19; (3) regardless of the respondents' genders, before COVID-19, older people were less likely to have a positive view; after the COVID-19 outbreak, they became more likely to have a positive view; and (4) married men became more likely to have positive view after COVID-19.

**Keywords**: View about women, Gender difference, Woman labor supply, COVID-19

**JEL Classification: J14; J16; J17**

1. **Introduction**

The Global Gender Gap Report (GGGR) rankings indicate that the amount of time spent by men on unpaid work (mainly domestic and volunteer work) is shorter than that spent by women in any countries (World Economic Forum, 2020). Even in countries where the gender gap is particularly low (i.e., Norway or the United States), women spend almost twice as much time as men on unpaid domestic work. In Japan, the share of time that women spend is more than four times that of men (World Economic Forum, 2020, p.11). According to the Global Gender Gap Index rankings, in 2020, Japan was the lowest among Group 7 countries and ranked 121st among 153 countries (World Economic Forum, 2020).[1] However, Japan's economic condition will rapidly change owing to the aging society. The labor shortage has gradually become apparent in recent years, which increases the demand for women's labor force to reduce the gap between demand and supply in the labor market.

In response to the COVID-19 pandemic, remote work has been adopted, which women are likely to do (Couch et al., 2022). The effects of remote work may differ by gender (e.g., Blazquez et al., 2023.; del Boca et al., 2022; Depalo & Pereda-Fernandez, 2023; Landivar et al., 2022). Berniell et al. (2023) found that working from home mitigated job losses, especially for women with children. Even after normalization of daily life, remote work is expected to reduce the gender gap in the labor market, especially in countries where the gender gap is large, such as Japan. Although a flexible work style was introduced, work may still be affected by gender identity (Akerlof & Kranton 2000); if so, women could not enjoy benefit of remote work (Augustine & Prickett, 2022; Fukai et al., 2023; Yamamura & Tsustui, 2021a, 2021b)[2].

In many countries, Bluedorn et al. (2023) found larger declines in women's employment rates compared to men during the pandemic period. COVID-19 had a larger impact on non-essential workers than on essential workers. Women essential workers, such as health care workers, are more likely to be influenced because people are more likely to demand women essential workers than men ones (Meekes et al., 2023). People might consider women essential workers to be more important than they were before the pandemic because they faced difficulty when health care service could not meet its demand under the emergency and unexpected situations such as lockdown. Further, in the drastically aging society of Japan, demand for health care workers is increasing. Questions arise: will the views about women's role and labor supply be persistent? How did COVID-19 influence gender identity and the view about women's active work? This paper aims to examine how subjective views about women's active work changed before and after COVID-19.

---

[1] G 7 countries consist of USA, UK, Canada, French, Italy, Germany, and Japan.
[2] In Italy, there is no gender differences about remote working during the lockdown to cope with COVID 19 (Bettin, et al. 2024).

We independently gathered individual-level panel data through Internet surveys to construct the panel dataset. An advantage of this study is that the data covered the period 2016−2024 and were valuable to explore the change of societal and economic conditions before and after COVID-19. Based on the data, we conducted a fixed effects (FE) estimation. The major findings are as follows: (1) people came to have a more positive view about women's active work; (2) change of the view was positively correlated with age if people experienced COVID-19, even though the view was negatively correlated with age before experiencing COVID-19. The strength of this study is that the panel dataset enabled us to utilize the natural experiment setting where unexpected exogenous shock occurred. Estimation results imply that people changed their view in response to labor market conditions by learning about the shortage of labor supply. This tendency is remarkable for older people, who are generally believed to be more conservative and to have negative views about women's work in societies where the gender gap is large.

This paper is organized as follows. The related literature is reviewed in Section 2, and the survey design and data are explained in Section 3. The empirical framework is described in Section 4. We present the estimation results and their interpretation in Section 5. Finally, Section 6 proposes concluding remarks.

## 2. Related Literature

Gender identity is one of key factors leading to the gender difference in time allocation for unpaid housework and paid work (Akerlof & Kranton, 2020; Yamamura & Tsutsui, 2021c). The gender gap of work style was one of major issues during COVID-19. In Canada, COVID-19 had a negative effect on women's employment and income, increasing gender inequalities (Ham, 2021; Singh et al., 2022). Lockdown, in the early phase of COVID-19, increased the gender gap of employment (Qian & Fuller, 2020), especially in the non-essential sector (Bettin et al., 2024; Blazquez et al, 2024). Meanwhile, in Spain, women's full-time employment did not decline significantly compared to that of men in 2020 after the onset of the COVID-19 pandemic (Villarreal & Yu, 2022).[3] COVID-19 has reduced the gender gap in employment in Indonesia (Halim et al., 2023) and in Turkey (Ilkkaracan & Memis 2021). In Australia, comparing to before COVID-19, the gender gap for housework has narrowed (Craig & Churchill, 2021).

Using data covering six countries after COVID-19, Dang and Viet (2021) found that women were 24% more likely to lose their job than men. During the lockdown period in China, compared to before COVID-19, on average, labor demand for women- and men-dominated industries decreased by 39% and 38%, respectively. On the labor supply side, the number of female and male job applicants decreased by 30% and 27%, respectively. This indicates that women search for jobs more actively than men (Hu et al., 2024).

Under the emergent situation when gender difference in labor supply is considered, a critical aspect is whether one has school-aged children. In various countries, in response to COVID-19, schools were closed, which increased the burden of parental childcare. School closing and childcare and their association to parental labor supply after the outbreak of COVID-19 have been widely analyzed.

---

[3] In Spain, COVID-19 led to an increase in the gender gap in total hours worked, not only in paid but also in unpaid work (Farre et al., 2022). Women were more likely to suffer job loss than men during the pandemic (Blazquez et al., 2023).

During COVID-19, women increased their time spent on childcare and housework in the UK (Oreffice & Quintana-Domeque, 2021) and Italy (Del Boca et al., 2020). In Japan, increased childcare responsibilities within a household resulted in a decline of mothers' employment (Fukai et al., 2023). In the USA in 2020, gender gaps in employment and hours worked for women with school-age children widened, while women were more inclined to do telework (Couch et al., 2022). In Argentina, during the lockdown, the load of men and women's paid work became more equitable, whereas women mainly burdened the additional housework and childcare, which increased the gender gap of unpaid work (Costoya et al., 2022); conversely, fathers with children spent about half of their daily hours helping children with schoolwork and reduced their earnings (Hupkau et al., 2023).

Schools' reopening increased labor supply, especially for single mothers (Beauregard et al., 2022). In particular, less mature children caused their own mothers to reduce their labor supply (Landivar et al., 2022; Meekes et al., 2023). Working parents without school-aged children did not experience a change in work hours and earnings, which implied that the school closure effect depended on the children's age (Heggeness, 2020; Kim et al., 2022). These studies have examined the short-term effect of COVID-19 on gender difference.

From a longer-term perspective, early school closures are considered to have had a negative effect on parental labor supply (Amuedo-Dorantes et al., 2023). The long-lasting impact of COVID-19 caused women's employment to move back toward pre-pandemic levels more rapidly than men's, whereas the gender gap in earnings widened in Australia (Risse, 2023). In Germany, owing to closures of daycare centers, gender inequality in the division of unpaid care work increased in the early stage of the pandemic. However, this tendency was temporary and not persistent (Jessen et al., 2022).

In the aging societies of developed countries, generations less likely to have school-aged children occupied a larger share of population compared to younger generations. For elderly European people in 2020, compared with men, women were more likely to increase the amount of worked hours or remote work compared to before COVID-19 (Depalo, D & Pereda-Fernandez, 2023).

3. **Data and Survey Method**

This study used the panel data gathered through Internet surveys covering 2016–2024. The initial survey was commissioned to Nikkei Research Company (NRC), which has know-how in Internet academic surveys through sufficient experience. Conducting surveys, the NRC continued to collect data until the target sample was sufficient.[4] Surveys were conducted six times in total: three times before the COVID-19 pandemic in 2016 (July 12−19), 2017 (July 11−19), and 2018 (Oct 3−11), and three times after the outbreak in 2021 (Dec 2−8), 2023 (Mar 2−9), and 2024 (Feb 1−9). Identical questionnaires were sent to the participants through the Internet to construct a panel dataset. However, as generally occurred in panel survey research, some participants dropped out as they stopped participating. Hence, to keep sample size, new participants were recruited through subsequent surveys. Accordingly, the sample structure became an unbalanced panel.

The key variable of this paper is the view about women's active work. Respondents were asked about it through the following question:

*"The government should create a society where women can fully demonstrate their abilities and play*

---

[4] In the survey, respondents were classified into 60 groups, with five cohorts (20−29, 30−39, 40−49, 50−59, and over 60), each with two genders. Accordingly, the groups were 10. Further, these groups were further divided into six regions commonly used in Japan (Hokkaido-Tohoku, Kanto, Chubu, Kinki, Shikoku-Chugoku, and Kyushu-Okinawa), which led to 60 categorized groups. In the surveys, observations were assigned to the group that was almost equivalent to the population share in each category of the Japan Census. Thus, the sample of this study can be considered representative of the population structure of Japan.

*an active role in the workplace."*

*Choose one of five choices from Strongly Agree (5) to Strongly disagree (1).*

This question was not included in the 2018 survey. Thus, the data used in this study did not include data gathered in 2018. Observations in the sample were 3,8766. Some respondents did not respond to some control variables such as marital status, jobs status, and household income levels. Accordingly, observations were reduced to 22,749 and used for the full-sample estimations reported in Table 2. We scrutinized the view about women's active work by dividing the sample into women and men. Finally, the number of observations in the sub-sample used in the estimation was reduced to 9,834 and 12,815 for women and men, respectively. These were used for sub-sample estimations in Tables 3 and 4.

Table 1 provides descriptions of key variables and their mean values and standard deviations. The sample used for Table 1 consists of 13,265 observations that are equivalent to those used for fixed effects ordered logit (FEOL) estimations reported in column (1) of Table 5. The sample is limited to those who changed their views during the studied period. However, values reported in Table 1 are almost equivalent to those in the sample including 22,749 observations. The mean of *Covid-19* is 0.44, which indicates that 44% observations were collected after the pandemic. We calculated the respondents' ages based on birth year. We considered how drastically increasing aging population influenced demand for women workers that would be expected to work in the health care sector. Therefore, we divided the sample into three groups; respondents younger than 50, those aged 50–64, and those aged over 65. Then, we considered differences in views about women's active work between age groups. The means of *Age5064* and *Age65* are 0.43 and 0.08, respectively. This indicates that respondents aged 50–64 and those aged over 65 occupied 43% and 8%, respectively. Accordingly, about 50% of respondents were younger than 50. Based on the information of job status, we made the dependent variable *Full Employ,* which is as dummy variable taking a value of

1 if respondents were employees, otherwise 0.[5]

Apart from the variables of Table 1, in estimations, we controlled basic characteristics such as household income and residential prefectures.

Figure 1 shows the changes in view about women's active work in the study period. Before COVID-19, men were more likely to have a more positive view about it than women, and the difference between men and women is statistically significant. Despite the particularly large gender inequality in the GGGR (World Economic Forum, 2020), men seemed to be progressive compared to women. This indicates that Japanese women tended to have a more traditional and conservative view compared to women of other countries. However, both men and women came to have a more positive view after COVID-19 compared to before it. Further, from 2017 to 2021, women drastically became more progressive in supporting their own active working compared to men; then, their views stabilized between 2021 and 2024. This implies that COVID-19 led Japanese people to demand women's active work. This partly reflects the rapidly aging society of Japan. However, the trend has persisted after 2015 for the following reasons. After WWII, births drastically increased. The large number of people born between 1947 and 1949 were called the "baby boom generation" (*Dankai no Sedai*) in Japan. People belonging to the baby boom generation retired to become elderly after 2010. Hence, the rate of population over 65 years of age increased from 24% to 30% from 2010 to 2024 (Appendix Figure A 1). However, the trend did not distinctly change after the outbreak of COVID-19. Therefore, the remarkable change in view about women's work reflects the impact of COVID-19 rather than that of the aging society.

Figures 2a, 2b, and 2c compare the distribution of women's work before and after COVID-19. Before

---

[5] In the questionnaire, full-time employees can be classified into various categories: company employee (full-time, non-managerial), company employee (managerial), company director/manager, contract worker, civil servants, teacher, heath care workers (doctor, nurse, therapist, etc.), and other professional (lawyer, accountant, tax accountant, etc.).

COVID-19, the rates of choosing 3 ("neutral") were the highest among five choices and were approximately 40% for women (Figure 2b) and 30% for men (Figure 2c). However, after COVID-19, the highest rate for 40% of women was 4 ("Agree") (Figure 2b). Meanwhile, for men, the rate of 4 increased from 24% to 33% and that of 5 from 18% to 22%, although the rates of choosing 3 ("neutral") continued to be the highest.

Considering the difference in cohorts, the eldest group of those born before 1955 shows the lowest values before COVID-19 and the highest ones after COVID-19 in Figures 3a, 3b, and 3c. The younger the group, the smaller the gaps between before and after COVID-19. As shown in Figure 3, for the women sample, changes across cohorts are not so large. In contrast, in Figure 3c, for the men sample, changes across cohorts are especially large. For the oldest group of men sample, the gap between periods is distinctly large. Then, for the youngest group of those born after 1996, we observe no statistical difference. In our interpretation, men of the older generation are less able to take care of themselves than women because they have hardly done house chores in their lives. Therefore, after experiencing COVID-19, elderly men were more likely to demand health care service provided by women compared to younger men.

## 4. Hypotheses and Methodology

### 4.1 Hypotheses

If unexpected events such as pandemics occur, women's employment is crucial to prepare for their husbands' job loss, and wives become the breadwinners for their families (Brini et al., 2024). Confronting the unexpected risk, on one hand, married men would like their wives to work more actively to cope with the risk that the men could lose their job. On the other hand, married women also recognize the importance of their own work for their families—for example, when their husbands cannot work. Experiencing COVID-19 led people to be positive about women's work. Hence, the following *Hypothesis 1* was proposed:

*Hypothesis 1: After COVID-19, people came to have a positive view about women's active work.*

Repeated pandemics put health care workers at risk of infection themselves. Shortages in health care services led to recognizing the importance of essential workers. In particular, elderly people need care services to live. Inevitably, the older a person becomes, the more important care services are. Therefore, elderly people's demand for women labors is larger than younger ones, assuming that women are more qualified for health care service than men. Thus, *Hypothesis 2* was raised:

*Hypothesis 2: Elderly people are likely to have a positive view about women's active work.*

### 4.2 Fixed Effects Model

To test *Hypotheses 1 and 2*, the estimated function takes the following form:

$$Work\ Women_{it} = a_1 Covid_t \times Age5064_{it} + a_2 Covid_t \times Age65_{it} + a_3 Age5064_{it} + a_4 Age65_{it} + a_5 Covid_t + XB + k_i + u_{it},$$

where *Work Women $_{it}$* is the dependent variable for individual $i$ at time $t$. The regression parameters and error term are denoted by $\alpha$ and $u$, respectively. $k_i$ is the time-invariant characteristic of the respondents to capture norm inherited from parents, genders, circumstances of their childhood, and educational background. Instead of cohorts that capture respondents' birth year, we simply include age groups. This is because the effect of cohorts depends on the survey year. To take an example, respondents aged 47 in 2016 became aged 55 in 2024. Here, we include the interaction terms between *Covid* and age group dummies to consider how and to what extent the effect of age changed before and after COVID-19. That is, we explore the effect of ages under different circumstances. Hence, we include dummies of age groups rather than cohort dummy. In our specification, the base age group includes those younger than 49. Hence, we include *Age5064 (Age65)* to show how individuals of older age groups have different views of women's work compared to those aged under 49. We used the FE model to control various time-invariant variables. X is the vector of control variables such as interaction terms $Covid_t \times Married_{it}$ and $Covid_t \times Full\ Worker_{it}$ and with Household income dummies. Further, B is the vector of their coefficient. The dependent variables are ordered variables; thus, the FEOL model is used as an alternative model.

The key variable is *Covid*, and its interaction terms $Covid \times Age5064$, and $Covid \times Age65$. From *Hypothesis 1*, the coefficient of *Covid* is predicted to have a positive sign. Generally, older generations are considered to have traditional views; thus, the predicted signs of *Age5064* and *Age65* are negative. However, as shown in *Hypothesis 2*, experiencing COVID-19 changed the view of elderly people. Hence, the signs of $Covid \times Age65$ and $Covid \times Age5064$ are predicted to be positive.

As observed in Figures 2 and 3, the view about women's active work differs by gender. Hence, for comparing the estimation results of men with those of women, apart from the full sample, subsamples of women and men are used for estimations. For women,

## 5. Results and Interpretations

Table 2(a) shows the full-sample estimation results of the FE model. As for key variables, we observe the positive sign of *Covid* and statistical significance at the 1% level. $Covid \times Age5064$, and $Covid \times Age65$ also show a positive sign and statistical significance at the 1% level. Further, the values of the coefficient of $Covid \times Age65$ are over two times larger than those of $Covid \times Age5064$. COVID-19 changed the view of retired people. As shown in the lower part of Table 2, the difference in coefficient between $Covid \times Age65$ and $Covid \times Age5064$ is statistically significant at the 1% level. Additionally, we observe a negative sign and statistical significance at the 1% level for *Age65* but not for *Age5064*. This implies that before COVID-19, retired aged people were less likely to support women's active work compared to those younger than 50. However, no significant difference exists between those aged 50–64 years and those younger than 50. These are consistent with *Hypotheses 1 and 2.*

Concerning control variables, the significant positive sign of *Covid× Married* reflects that the experience of the pandemic led married persons to be positive about women's active work to maintain the household's income level when unexpected events occur. The significant positive sign of *Full-worker* implies that, before COVID-19, full-time workers were more likely to be positive about women's active work. However, non-full-time workers changed their view more drastically than full-time workers after the pandemic, which led *Covid × Full-worker* to have a significant negative sign.

Turning to the results of sub-sample estimations, Tables 3 and 4 report the results using women and men samples, respectively. As for key variables, their results in Tables 3 and 4 are almost the same as those in Table 2. Hence, these *Hypotheses 1* and *2* are supported regardless of gender. However, the control variables exhibit gender differences. For women in Table 3, we do not observe statistical significance in any results of *Full-worker, Covid × Full-worker,* and *Age65*; that is, for women, the view about women's active work does not depend on their work status and age. Meanwhile, for men in Table 4, the results of these variables are similar to those in Table 2. The significant sign of *Covid× Married* is negative for women but and positive for men. In our interpretation about the negative sign for women compared to married women, unmarried women were more likely to recognize the importance of their work when they faced difficulty during the pandemic because they could not depend on their husbands' income.

As for the robustness check, the results of FEOL are reported in Table 5. These results of key variables are almost equivalent to what we observed in Tables 2–4. Combined, the results of Tables 2–5 strongly support *Hypotheses 1* and *2.* However, concerning control variables in sub-sample estimations, the statistical significance of their coefficients disappears, except for the statistical positive sign of *Covid× Married* using the men sample. In FEOL, coefficients cannot be interpreted as suggesting the variables' size of effect. To consider the size of effect, the marginal effect should be considered. However, in FEOL, we can obtain five different marginal effects according to the probability of respondents choosing *View*

*Woman Work*. For instance, we can calculate the five marginal effects of *Covid × Age65* when respondents strongly agree, agree, neutral, disagree, and strongly disagree. In Table 5, we report the marginal effects of *Covid × Age65* and *Covid × Age5064* on the probability that respondents strongly agree with women's active work. The marginal effects of *Covid × Age65* are 1.86 and 0.143 for women and men, respectively. Women became more likely to choose "strongly agree" with their active working by 18.6% after COVID-19 compared to before it, whereas men became more likely to choose "strongly agree" by 14.3% after COVID-19 than before it. Hence, the effect of COVID-19 is about 4% larger for women than for men. In our interpretation, elderly women can more easily communicate with women health care workers to explain their need compared to men because men health care workers are seemingly less able to understand women's condition. In other words, women health care workers are more able to meet the requests of not only elderly women but also elderly men.

Further, regardless of gender, the marginal effect of *Covid × Age65* is larger than that of *Covid × Age5064*. Women and men became more likely to choose "agree" by 10.6% and 6.7% points, respectively. Considering *Covid × Age65* and *Covid × Age5064* together, the older the persons, the more they came to be positive about women's active work. This reflects that women's labor is required more by the health care sector, which is more important for older people.

## 6. Conclusions

Among developed countries, women's social status in Japan is the lowest, with a large gender gap in the labor market and political participation. The contribution of this study is that it utilized a natural experiment setting to provide evidence that unexpected exogenous shock due to the pandemic triggered the change in view about women's active work in a society where the gender gap is large. This study showed that men's view about women's active work is more positive than women's own view. Regardless of gender, people have become more positive about women's active work. Particularly, women's view about their active work became more positive than that of men. Regardless of gender, older people were less positive before COVID-19 but became more positive than younger people. Married men become more likely to have a positive view after COVID-19 than before it.

The findings imply that experiencing the pandemic led people to consider health care workers as essential. Especially, senior people became aware of the importance of health care services because they are more likely to demand health care support compared to younger people. Generally, women are considered to feel a real sense of life and favor the attention to detail required for health care. Further, the rapid aging of society has led to a shortage of workers, and senior people's view about women's active work to fill the labor shortage has drastically changed to positive.

Nevertheless, in addition to the factors we considered in this study, various other factors influence the change in view about women's active work. The aging society naturally led to an increase in demand for health care workers and thus women's active work, regardless of COVID-19. Owing to limitation of the data, we could not control the effect of the aging society. Further, government policy also influenced this view. The work environment has changed, and work styles have diversified—for example, through remote work. The influx of foreign labor has also increased. Further, exploring to what extent the findings of this paper can be generalized in other developed countries with rapidly aging populations is necessary. These issues should be addressed in future studies.

**Funding Information.** This work was supported by the Japan Society for the Promotion of Science (JSPS) under Grants 16H03628, 20H05632 and 25H00388, .

*Household,* 19(1), 41-60.

Yamamura, E., and Tsutsui., Y. (2021b). School closures and mental health during the COVID-19 pandemic in Japan. *Journal of Population Economics,* 34(4), 1261-1298.

Yamamura, E., and Tsutsui, Y. (2021c). Spousal gap in age and identity and their impact on the allocation of housework. *Empirical Economics,* 60(2), 1059−1083.

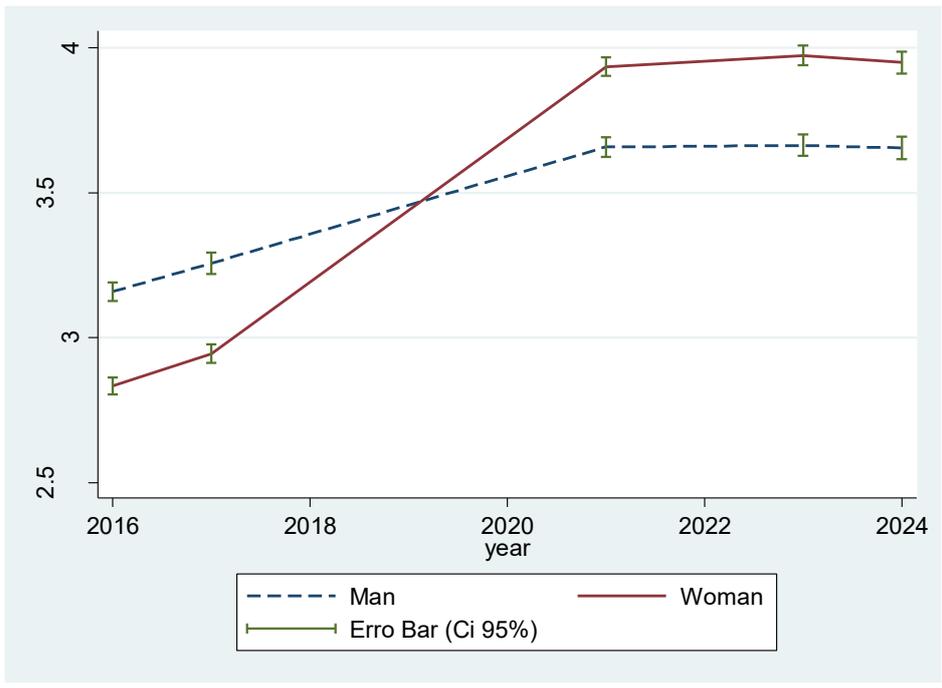

**Figure 1.** View about women's work: Men versus women

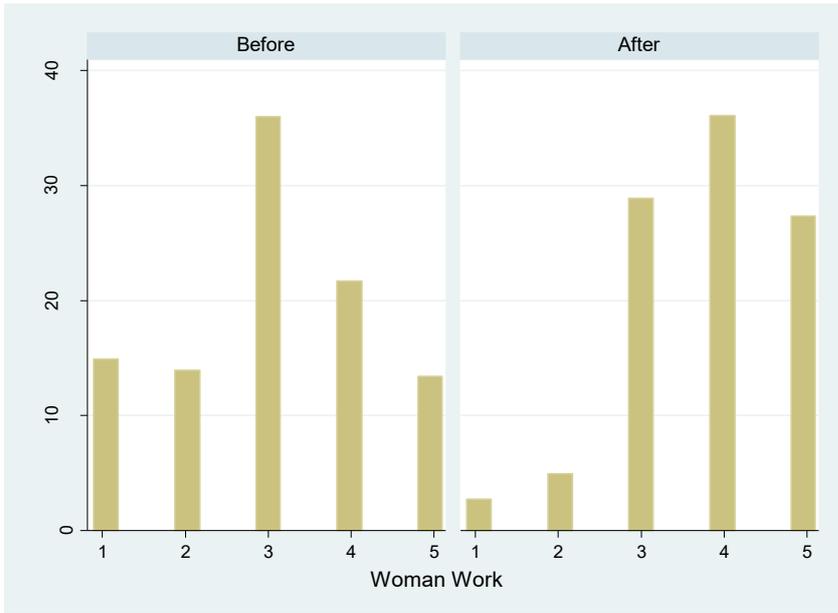

**Figure 2a.   Full sample: Comparing histogram of women's work before and after COVID-19**

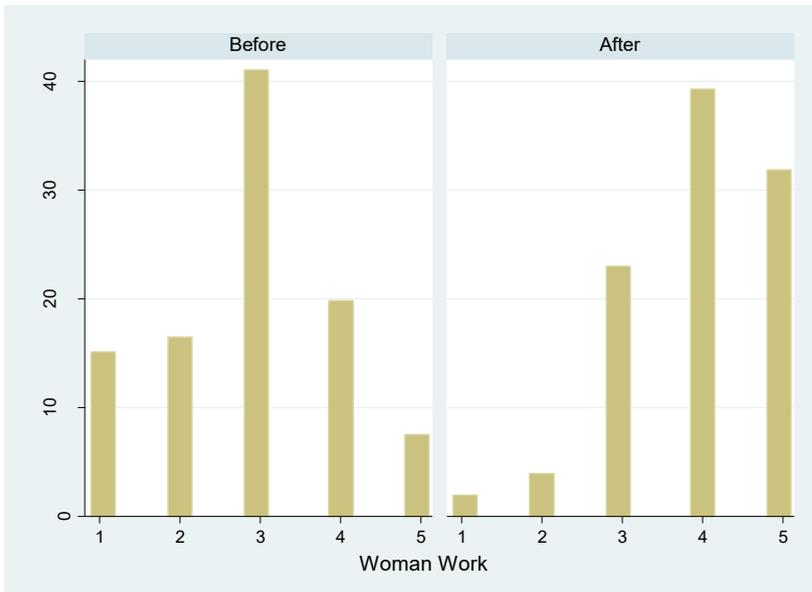

**Figure 2b.   Woman sample: Comparing histogram of women's work before and after COVID-19**

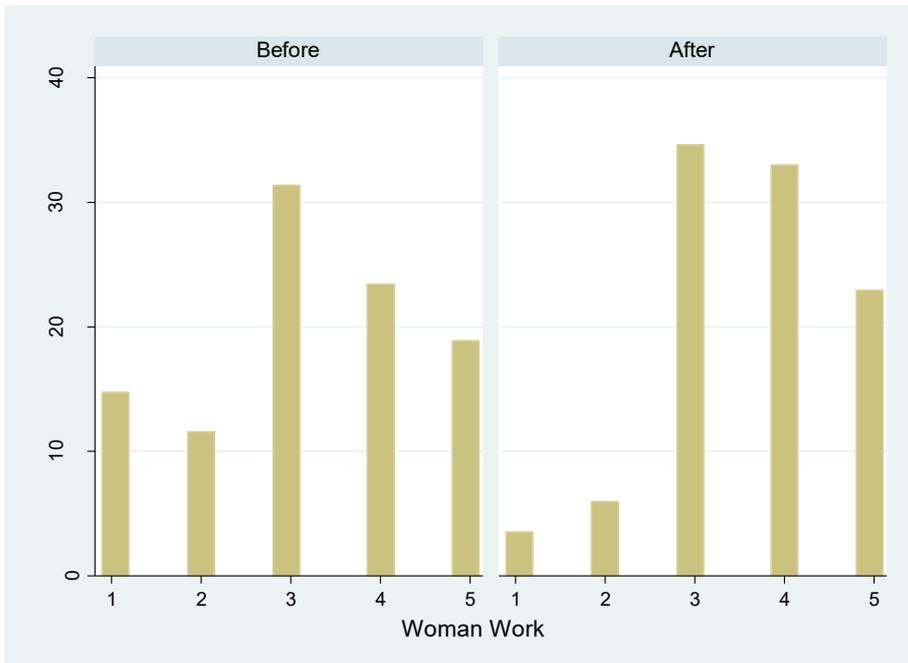

**Figure 2c.　Man sample: Comparing histogram of women's work before and after COVID-19**

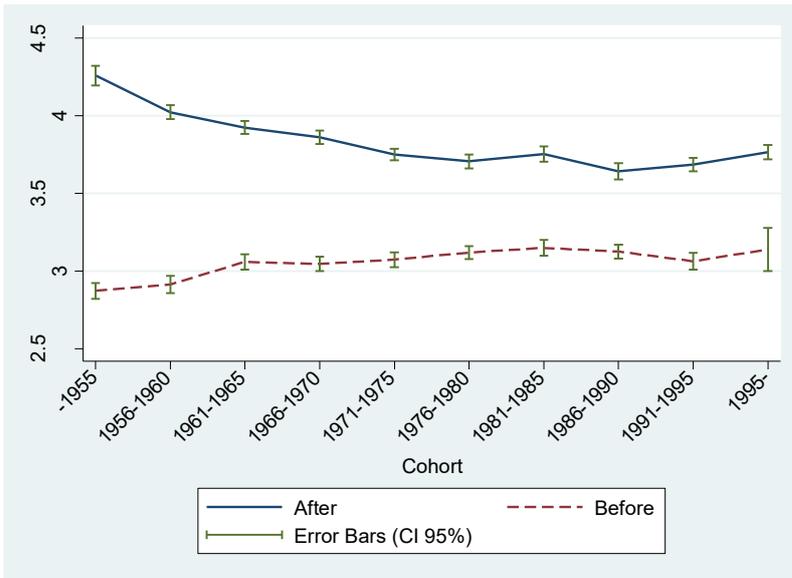

**Figure 3a. Full sample: Comparing women's work before and after COVID-19 by cohort**

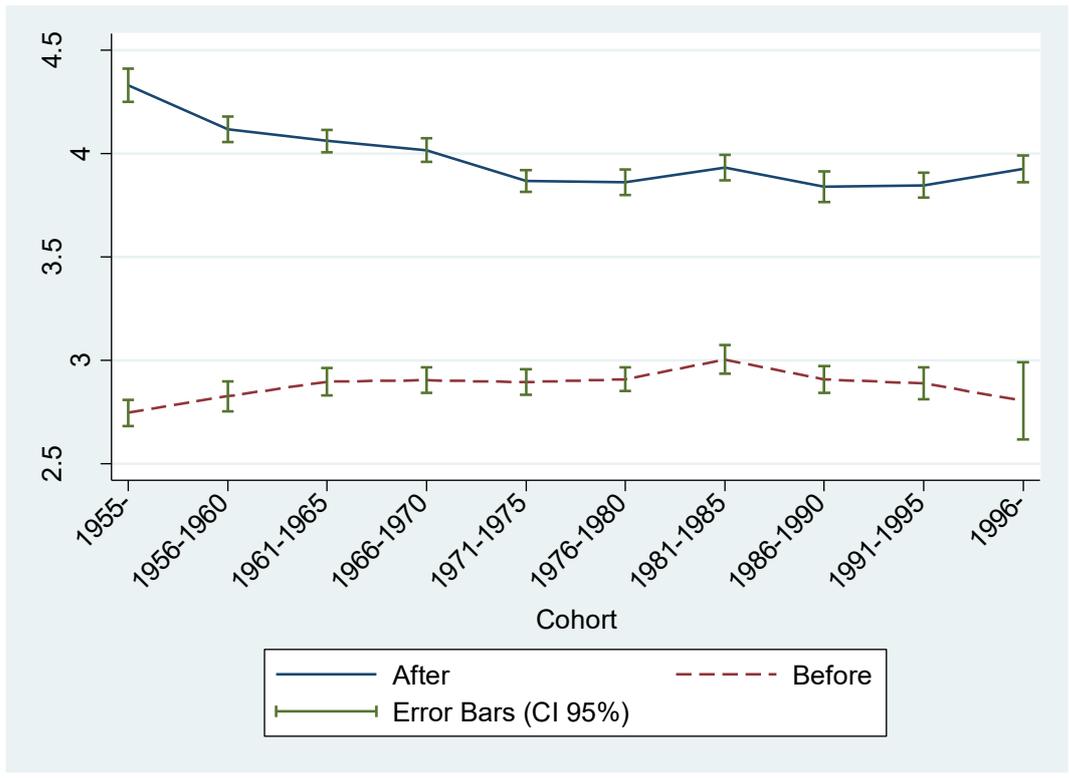

**Figure 3b. Woman sample: Comparing women's work before and after COVID-19 by cohort**

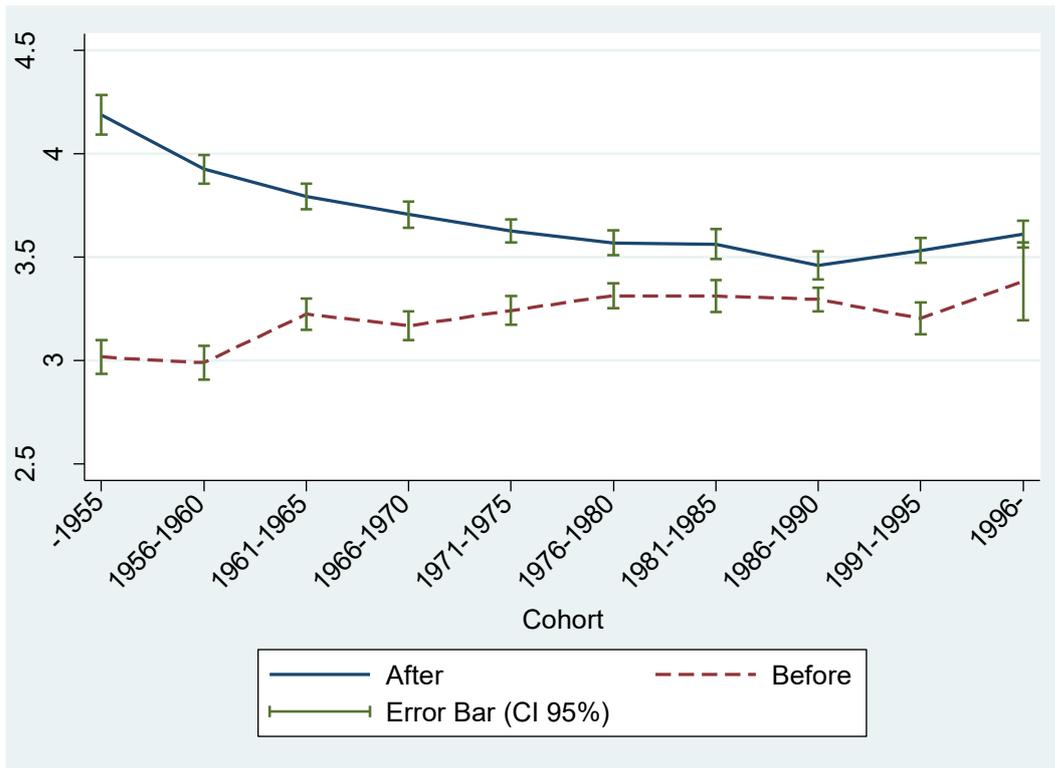

**Figure 3c. Man sample: Comparing women's work before and after COVID-19 by cohort**

Table 1. Definitions of variables and their basic statistics.

| Variables | Definition | Mean | s.d. |
|---|---|---|---|
| *View Woman Work* | Question. The government should create a society where women can fully demonstrate their abilities and play an active role in the workplace. Strongly Agree (5) - Strongly disagree (1) | 3.38 | 1.28 |
| *Covid* | Equals 1 for the survey conducted in 2021, 2023, or 2024, while being 0 for the survey in 2016 and 2017. | 0.44 | 0.49 |
| *Age5064* | Equals 1 if respondent's ages are between 50 and 64, otherwise 0. | 0.43 | 0.49 |
| *Age65* | Equals 1 if respondent's ages are equal or over 65, otherwise 0. | 0.08 | 0.37 |
| *Married* | Equals 1 if respondent is married, otherwise 0. | 0.63 | 0.48 |
| *Full-worker* | Equals 1 if respondent is a full-time worker, otherwise 0. | 0.56 | 0.49 |

Note: Sample of women and men in Table 5 is used; observations are 13,265.

Table 2. Estimation results: dependent variable of *View Woman Work*. Fixed Effects model (FE). Sample of women and men.

|  | (1) | (2) |
|---|---|---|
| *Covid ×Age5064* | 0.296*** | 0.272*** |
|  | (0.066) | (0.065) |
| *Covid ×Age65* | 0.691*** | 0.593*** |
|  | (0.103) | (0.104) |
| *Covid ×Married* |  | 0.135** |
|  |  | (0.062) |
| *Covid ×Full-worker* |  | −0.271*** |
|  |  | (0.057) |
| *Age5064* | −0.042 | −0.100 |
|  | (0.026) | (0.062) |
| *Age65* | −0.066*** | −0.065*** |
|  | (0.023) | (0.093) |
| *Covid* | 0.482*** | 0.644*** |
|  | (0.052) | (0.062) |
| *Married* | −0.062 | −0.047 |
|  | (0.082) | (0.083) |
| *Full-worker* |  | 0.181*** |
|  |  | (0.061) |
| Difference test *Covid ×Age5064* − *Covid ×Age65-* | F-16.1 Prob>F=0.00 | F-10.1 Prob>F=0.00 |
| Within R-square | 0.18 | 0.17 |
| Obs. | 22,649 | 22,649 |

Note: In all columns, 13 household income dummies and 46 residential prefecture dummies are included; however, the results are not reported. Numbers within parentheses are robust standard errors clustered by individuals. ***, **, and * indicate statistical significance at the 1%, 5%, and 10% levels, respectively.

Table 3. Estimation results: dependent variable of *View Woman Work*. Fixed Effects model (FE). Sample of women

|  | (1) | (2) |
|---|---|---|
| *Covid ×Age5064* | 0.347*** (0.094) | 0.349*** (0.095) |
| *Covid ×Age65* | 0.672*** (0.142) | 0.693*** (0.145) |
| *Covid ×Married* |  | −0.168* (0.095) |
| *Covid ×Full-worker* |  | 0.029 (0.091) |
| *Age5064* | −0.072 (0.088) | −0.073 (0.088) |
| *Age65* | −0.041 (0.128) | −0.050 (0.127) |
| *Covid* | 1.020*** (0.086) | 1.000*** (0.103) |
| *Married* | 0.017 (0.129) | −0.005 (0.128) |
| *Full-worker* |  | 0.001 (0.089) |
| Difference test *Covid ×Age5064 Covid ×Age65-* | F-5.81 Prob>F=0.01 | F-6.39 Prob>F=0.01 |
| Within R-square | 0.33 | 0.32 |
| Obs. | 9,834 | 9,834 |

Note: In all columns, 13 household income dummies and 46 residential prefecture dummies are included; however, the results are not reported. Numbers within parentheses are robust standard errors clustered by individuals. ***, **, and * indicate statistical significance at the 1%, 5%, and 10% levels, respectively.

Table 4. Estimation results: dependent variable of *View Woman Work.* Fixed Effects model (FE). Sample of men

|  | (1) | (2) |
|---|---|---|
| *Covid ×Age5064* | 0.218** (0.087) | 0.218** (0.086) |
| *Covid ×Age65* | 0.621*** (0.148) | 0.550*** (0.150) |
| *Covid ×Married* |  | 0.257*** (0.086) |
| *Covid ×Full-worker* |  | −0.184** (0.093) |
| *Age5064* | −0.076 (0.084) | −0.076 (0.083) |
| *Age65* | −0.019 (0.135) | −0.009 (0.134) |
| *Covid* | 0.239*** (0.063) | 0.343*** (0.085) |
| *Married* | −0.007 (0.107) | −0.026 (0.107) |
| *Full-worker* |  | 0.134 (0.089) |
| Difference test *Covid ×Age5064 Covid ×Age65-* | F-8,17 Prob>F=0.00 | F-5.47 Prob>F=0.02 |
| Within R-square | 0.10 | 0.10 |
| Obs. | 12,815 | 12,815 |

Note: In all columns, 13 household income dummies and 46 residential prefecture dummies are included; however, the results are not reported. Numbers within parentheses are robust standard errors clustered by individuals. ***, **, and * indicate statistical significance at the 1%, 5%, and 10% levels, respectively.

Table 5. Estimation results: dependent variable of *View Woman Work.* Fixed Effects Ordered Logit model (FEOL).

|  | (1)<br>*Women and men* | (2)<br>*Women* | (3)<br>*Men* |
|---|---|---|---|
| *Covid*<br>×*Age5064* | 0.469***<br>(0.143) | 0.705***<br>(0.232) | 0.404**<br>(0.184) |
| *Covid*<br>×*Age65* | 0.875***<br>(0.271) | 1.239***<br>(0.427) | 0.861**<br>(0.387) |
| *Covid*<br>×*Married* | 0.307**<br>(0.132) | −0.244<br>(0.247) | 0.478***<br>(0.176) |
| *Covid*<br>×*Full-worker* | −0.565***<br>(0.127) | 0.100<br>(0.227) | −0.313<br>(0.194) |
| *Age5064* | −0.165<br>(0.138) | −0.024<br>(0.235) | −0.155<br>(0.174) |
| *Age65* | 0.075<br>(0.265) | 0.189<br>(0.410) | 0.131<br>(0.370) |
| *Covid* | 1.373***<br>(0.140 | 2.127***<br>(0.261) | 0.712***<br>(0.180) |
| *Married* | −0.103<br>(0.190) | −0.041<br>(0.307) | −0.044<br>(0.247) |
| *Full-worker* | 0.350**<br>(0.145) | −0.069<br>(0.272) | 0.217<br>(0.193) |
|  | *Marginal effect*<br>*Prob (View Woman Work=5)* | | |
| *Covid*<br>×*Age5064* | 0.075***<br>(0.023) | 0.106***<br>(0.034) | 0.067**<br>(0.030) |
| *Covid*<br>×*Age65* | 0.140***<br>(0.043) | 0.186***<br>(0.064) | 0.143**<br>(0.063) |
| Pseudo R-square | 0.19 | 0.37 | 0.11 |
| Obs. | 13,265 | 5,413 | 7852 |

Note: In all columns, 13 household income dummies and 46 residential prefecture dummies are included; however, the results are not reported. Numbers within parentheses are robust standard errors clustered by individuals. ***, **, and * indicate statistical significance at the 1%, 5%, and 10% levels, respectively.

Appendix.

Figure A1.

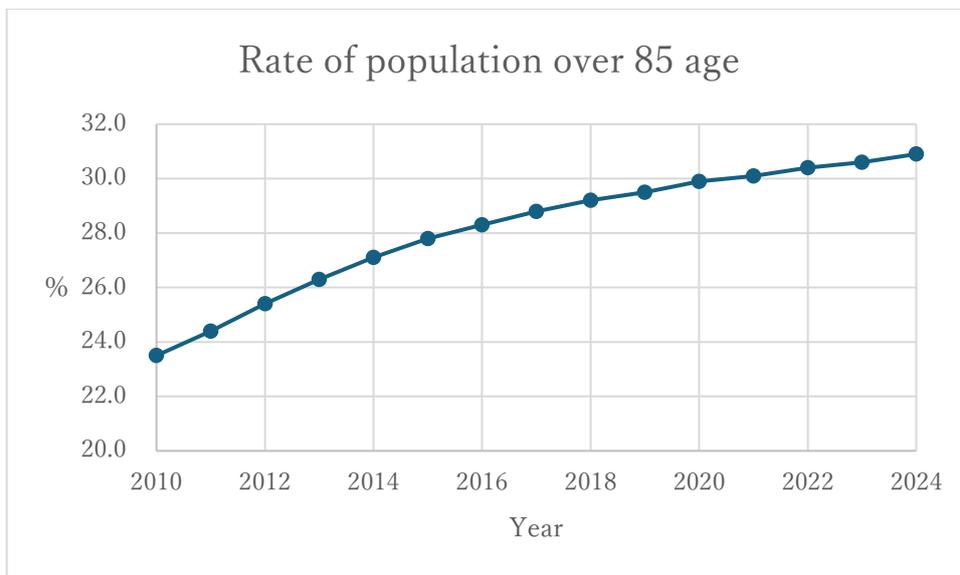

Source: Population Estimates for the Future of Japan (December 2006 Estimates) by National Institute of Population and Social Security Research.

https://www.ipss.go.jp/pp-newest/j/newest03/h3_1.html (Accessed on Feb 1, 2025)